# **Phase formation in CrFeCoNi nitride thin films**

Smita G. Rao[†*1], Boburjon Mukhamedov[†*2], Gyula Nagy[4], Eric N. Tseng[1], Rui Shu[1], Robert Boyd[3], Daniel Primetzhofer[4], Per O. Å. Persson[1], Björn Alling[2], Igor A. Abrikosov[2], Arnaud le Febvrier[1], Per Eklund[1]

[1] *Thin Film Physics Division, Department of Physics, Chemistry, and Biology (IFM), Linköping University, Linköping, Sweden-58183.*

[2] *Theoretical Physics Division, Department of Physics, Chemistry, and Biology (IFM), Linköping University, Linköping, Sweden-58183.*

[3] *Plasma & Coatings Physics Division, Department of Physics, Chemistry, and Biology (IFM), Linköping University, Linköping, Sweden-58183.*

[4] *Department of Physics and Astronomy, Uppsala University, Lägerhyddsvägen 1, S-75120 Uppsala, Sweden*

[†]*Corresponding author: e-mail address: smita.gangaprasad.rao@liu.se; boburjon.mukhamedov@liu.se*

## **Abstract**

As a single-phase alloy, CrFeCoNi is a face centered cubic (fcc) material related to the archetypical 'high-entropy' Cantor alloy CrFeCoNiMn. For thin films, CrFeCoNi of approximately equimolar composition tends to assume an fcc structure when grown at room temperature by magnetron sputtering. However, the single-phase solid solution state is typically not achieved for thin films grown at higher temperatures. The same holds true for Cantor alloy-based ceramics (nitrides and oxides), where phase formation is extremely sensitive to process parameters such as the amount of reactive gas. This study combines theoretical and experimental methods to understand the phase formation in nitrogen-containing CrFeCoNi thin films. Density functional theory calculations considering three competing phases (CrN, Fe-Ni and Co) show that the free energy of mixing, $\Delta G$ of $(CrFeCoNi)_{1-x}N_x$ solid solutions has a maximum at $x = 0.20$-$0.25$, and $\Delta G$ becomes lower when $x < 0.20$ and $x > 0.25$. Thin films of $(CrFeCoNi)_{1-x}N_x$ ($0.14 \geq x \leq 0.41$) grown by magnetron sputtering show stabilization of the metallic fcc when $x \leq 0.22$ and the stabilization of the NaCl B1 structure when $x > 0.33$, consistent with the theoretical prediction. In contrast, films with intermediate amounts of nitrogen ($x = 0.22$) grown at higher temperatures show segregation into multiple phases of CrN, Fe-Ni-rich and Co. These results offer an explanation for the requirement of kinetically limited growth conditions at low temperature for obtaining single-phase CrFeCoNi Cantor-like nitrogen-containing thin films and are of importance for understanding the phase-formation mechanisms in multicomponent ceramics.

---

[*] S.G.R and B. M. contributed equally to this study

# 1. Introduction

Multicomponent and high entropy alloys (HEA) were suggested in the 1980s [1,2] but gained attention in 2004 when two separate studies by Cantor et al., (2004) [3] and Yeh et al., (2004) [4] showed that alloys with five or more elements in nearly equal ratios can form common crystal structures. The Cantor alloy (CrMnFeCoNi) is among the most-studied high-entropy alloys and exhibits a face centered cubic (fcc) solid solution structure. At the time, it was theorized that the high mixing entropy ($\Delta S_{mix}$) resulted in a low Gibbs free energy ($\Delta G$) which in turn helped in the formation of single-phase solid solutions. This reasoning, however, does not hold true for all cases, especially when the realm of Cantor alloy-based materials has expanded into new directions [5][6]. Thus, the mechanisms of phase formation and stability in multicomponent alloys remain an important research topic.

Phase prediction in Cantor-based alloys has come a long way since Zhang et al., (2008) proposed that the atomic size difference played a role in determining the relative stability phase [7]. The stabilization of a bcc structure with additions of Al into the fcc Cantor alloy is one such example where this theory of atomic size difference comes into play [8,9]. Calculating the valence electron concentration (VEC) is another way to predict stable phases, the concept being that a VEC of $\geq 8$ suggests the stabilization of the fcc structure while a VEC of $\leq 6.8$ would indicate stabilization of the bcc structure [9]. This understanding is derived from the Hume-Rothery interpretation of VEC, its influence in orbital hybridization and metallic bonding [10]. These methods are relatively straightforward for equiatomic metallic bulk alloys of the 3d and 4d transition metals. However, these methods of phase prediction fail in more complex cases. For example, the VEC rule is valid only if one neglects the presence of multiple phases and irregularities in the composition [11–14]. For ceramic multicomponent systems, one must also take into consideration the position of the non-metal atoms in the lattice and the vacancies in the crystal structure [15].

Thin films introduce another level of complexity. Physical vapor deposition (PVD) techniques such as magnetron sputtering, condense material far from thermodynamic equilibrium with limited diffusion, and thus often allow the stabilization of metastable phases, in particular disordered solid solutions. In this work, we combine theoretical and experimental investigations to advance the understanding of the phase formation in thin films of the CrFeCoNi system, a simplified model of the full Cantor alloy. In addition to the metallic system, we consider more complex nitrogen containing multicomponent thin films, addressing the fundamental question of phase formation in the nitrogen-supersaturated and stoichiometric nitride films.

# 2. Methods

## 2.1 Theoretical methods

Electronic structure calculations of Cr-Fe-Co-Ni alloys with addition of interstitial nitrogen were carried out using the projector augmented-wave method [16] implemented in the VASP software [17,18]. Chemical disorder of solid solutions was described using special quasirandom structure (SQS) method [19]. All calculations were performed for 3x3x3 face centered cubic supercells (108 metal atoms) with equimolar concentration of metal atoms, i.e. $Fe_{0.25}Cr_{0.25}Co_{0.25}Ni_{0.25}$. To introduce nitrogen in $Fe_{0.25}Cr_{0.25}Co_{0.25}Ni_{0.25}$ solid solutions, we considered all octahedral interstitial positions of supercell as a separate "nitrogen-vacancy" sublattice that does not intermix with metal sublattice. By a vacancy, we mean an empty octahedral position. The amount of N atoms varied depending on its concentration in $Fe_{0.25}Cr_{0.25}Co_{0.25}Ni_{0.25}$ solid solution. We generated the $(CrFeCoNi)_{1-x}N_x$ SQSs with six different concentrations of N (mol): $x = 0$, 0.1, 0.2 0.25, 0.35 and 0.50, where $x = 0$ was a metal alloy without any nitrogen and $x = 0.50$ corresponded to a stochiometric nitride with B1 underlying crystal lattice. The ferrimagnetic and paramagnetic cases were considered for the calculations. Magnetic properties were accounted for within a model of collinear local magnetic moments. The magnitudes and orientations of the local moments were calculated self-consistently. The paramagnetic state of $Cr_{0.25}Fe_{0.25}Co_{0.25}Ni_{0.25}$ solid solution was described within a disordered local moment (DLM) picture [20] implemented in the framework of magnetic SQS method [21]. In the DLM, one can simulate paramagnetic $Cr_{0.25}Fe_{0.25}Co_{0.25}Ni_{0.25}$ as an effective eight-component alloy $(Cr\uparrow,Cr\downarrow)_{0.25}(Fe\uparrow,Fe\downarrow)_{0.25}(Co\uparrow,Co\downarrow)_{0.25}(Ni\uparrow,Ni\downarrow)_{0.25}$ with equal compositions of spin-up and spin-down atoms for each chemically non-equivalent alloy component. The generalized gradient approximation [22] was used to describe the exchange and correlation effects. The cutoff energy for plane waves was set to 450 eV. The convergence criterion for electronic subsystem was set to $10^{-4}$ eV for subsequent iterations. The full relaxation of SQSs was carried out, which includes the relaxation of supercell volume, shape, and ionic positions. The relaxation of ionic positions was carried out by calculating the Hellman-Feynman forces [23] and the stress tensor and using them in the conjugated gradient method. Relaxation was completed when the forces on the ions became of the order of $10^{-2}$ eV/Å.

To predict the phase stability of $(CrFeCoNi)_{1-x}N_x$ solid solutions, we carried out additional calculations of three competing phases, CrN-rich, Fe-Ni-rich and Co-rich. The compositions for reference states were chosen according to experimental data for films containing $x \approx 0.20$ - 0.22, which undergo segregation as discussed below in section 3.2.2. We tested this segregation trend for all N concentrations in $(CrFeCoNi)_{1-x}N_x$ SQSs, even though in the experiments there is no segregation in the films with $x$ lower and higher than ~ 0.20 - 0.22. The details of calculated structures for three competing phases, CrN-rich, Fe-Ni-rich and Co-rich, are available in supplementary information (see Table S2).

The mixing free energy of $(CrFeCoNi)_{1-x}N_x$ solid solutions is defined as:

$$\Delta G = \Delta E - T\Delta S_{conf} \qquad (1)$$

where $\Delta E$ and $\Delta S$ denote mixing total energy and mixing entropy which are defined as:

$$\Delta E = E_{(FeCrCoNi)N} - c_{CrN} \cdot E_{CrN} - c_{FeNi} \cdot E_{FeNi} - c_{Co} \cdot E_{Co} \qquad (2)$$

$$\Delta S = S_{(FeCrCoNi)N} - c_{CrN} \cdot S_{CrN} - c_{FeNi} \cdot S_{FeNi} - c_{Co} \cdot S_{Co} \qquad (3)$$

where $c_{CrN}$, $c_{FeNi}$ and $c_{Co}$ denote molar fractures of CrN-rich, FeNi-rich and Co-rich reference phases; $E_{(FeCrCoNi)N}$, $E_{CrN}$, $E_{FeNi}$ and $E_{Co}$ denote the ground state total energies of $(CrFeCoNi)_{1-x}N_x$ solid solution and three reference phases; and $S_{(CrFeCoNi)N}$, $S_{CrN}$, $S_{FeNi}$ and $S_{Co}$ denote their entropies. In this work, $c_{CrN} = 0.25$, $c_{FeNi} = 0.5$ and $c_{Co} = 0.25$. Since the composition of metals in (CrFeCoNi)$_{1-x}$N$_x$ solid solution is equimolar, there is an equal proportion of Fe and Ni in Fe-Ni-rich competing phase.

The entropy of the (CrFeCoNi)$_{1-x}$N$_x$ solid solutions and corresponding competing phases was calculated as a sum of configurational $S_{conf}$ and magnetic $S_{mag}$ entropies: $S = S_{conf} + S_{mag}$. Both $S_{conf}$ and $S_{mag}$ have been determined within mean-field approximation. The magnetic entropy is non-zero only for paramagnetic alloys, and is defined as $S_{mag} = k_B \sum_i c_i \cdot ln(1 + |m_i|)$ [24,25], where $m_i$ is the average magnetic moment of $i$-th component in the DLM state, and $c_i$ is a concentration $i$-th component. One should note that configurational entropy in nitrogen containing solid solutions has two contributions: from the disorder in metal sublattice and from the disorder in nitrogen sublattice. More details of calculations for $S_{conf}$ and $S_{mag}$ are available in Supplementary information (see section S1).

## 2.2 Experimental methods

CrFeCoNi metallic and nitrogen-containing thin films were grown on Si(100) substrates by magnetron sputtering from Cr$_{24}$Fe$_{32}$Co$_{24}$Ni$_{20}$ compound targets (provided by Plansee, Composite Materials GmbH, target thickness – 3 mm, 50.8 mm diameter) in an ultrahigh vacuum (base pressure $< 9 \times 10^{-7}$ Pa) magnetron sputtering system, described elsewhere [26]. Prior to deposition, the substrates were cleaned with acetone and isopropanol for 10 minutes in an ultrasonic bath and blow-dried with nitrogen gas. The compound target was operated at constant DC power at 100 W (4.9 W/cm$^2$). Depositions were carried out in an Ar+N$_2$ atmosphere where the working pressure was set at 0.53 Pa (4.0 mTorr) and the nitrogen content in the films was increased by changing the nitrogen flow ratio for different depositions from 10 % to 70 % (N$_2$/(Ar+N$_2$)).

The composition of the films grown was determined by a combination of ion beam analysis techniques. The nitrogen content relative to the cumulative metal content (N/metal) was obtained by time of flight-elastic recoil detection analysis (ToF-ERDA). Particle induced X-ray emission (PIXE) was used to resolve the concentration of individual metals (Cr, Fe, Co, Ni). In addition, Rutherford Backscattering Spectrometry (RBS) was used to determine the film thicknesses. Measurements were carried at the 5 MV NEC-5SDH-2 Pelletron Tandem accelerator of the Tandem Laboratory, Uppsala University, Sweden.

ToF_ERDA measurements were carried out using a 36 MeV 127I8+ beam. The beam incidence angle on the target was 67.5°, while the gas ionization chamber detector was placed at 45°. Depth profiles of the elemental composition were acquired from ToF-ERDA time and energy coincidence spectra with the Potku 2.0 software package [27].

PIXE and RBS measurements were carried out simultaneously, using a beam of 2 MeV He$^+$ primary ions. The incident angle between the beam and the sample normal was 5°, while the detection angle for RBS was 170° and for PIXE it was 135°. The acquired RBS spectra were evaluated using SIMNRA [28], the PIXE spectra were fitted using the GUPIX code.

X-ray diffraction (XRD) θ-2θ diffractograms were acquired in a PANalytical X'Pert PRO diffractometer operated using Cu-Kα radiation (λ = 1.54060 Å) at a voltage of 45 keV and current of 40 mA. All X-ray diffractograms were obtained at room temperature.

Transmission electron microscopy (TEM) and high-resolution TEM (HR-TEM) were carried out on select samples using a FEI Tecnai G2 TF 20 UT instrument operated at 200 kV. The cross-sectional TEM specimens were prepared by manual polishing down to a thickness of 60 μm, followed by Ar+ ion milling at 5 keV, with a 6° incidence angle, on both sides while rotating the sample in a Gatan precision ion polishing system.

Electron energy loss spectroscopy (EELS) characterization was performed using the Linköping double Cs-corrected FEI Titan3 60-300, operated at 300 kV. Elemental distribution maps were extracted from EELS spectrum images by background subtraction, using a power law, and choosing characteristic edges, N K-edge (402ev), Cr L3,2-edge (575, 584eV), Fe L3,2-edge (708, 721eV), Cr 3,2-edge (779, 794eV) and Ni L3,2-edge (855, 872eV) energy loss integration windows.

A scanning electron microscope (SEM), Leo 1550 Gemini, was used for obtaining surface images using the in-lens detector with an acceleration voltage of 8 keV. Backscattered SEM images were acquired using with a Zeiss Gemini 560 instrument fitted with an inlens detector and acquired using an acceleration voltage of 1 kV.

## 3. Results

### 3.1 DFT results

The mixing free energy of $(CrFeCoNi)_{1-x}N_x$ solid solutions was calculated with reference to three competing phases: CrN-rich, Fe-Ni-rich and Co-rich. We considered the isostructural segregation of $(CrFeCoNi)_{1-x}N_x$ solid solutions, which means that the Fe-Ni-rich and Co-rich reference phases have fcc lattice and CrN-phase has B1 lattice. The hcp lattice for Co was not considered for two reasons: (i) the Co-rich phase may contain iron which stabilizes the fcc lattice of Co [29] and (ii) all the other competing phases and $(CrFeCoNi)_{1-x}N_x$ solid solution have fcc or B1 structure. Table S2 in Supplementary information shows the number of atoms in the calculated SQS supercells for equimolar $(CrFeCoNi)_{1-x}N_x$ solid solutions and three reference phases.

Table 1 shows the chemical compositions of competing phases corresponding to different concentrations of N in the material system CrFeCoNi-N. The occupancy of nitrogen in competing phases was prioritized in the following order: first CrN, then Fe-Ni-phase and last Co-phase. This N distribution among the competing phases was based on the formation energies of mononitrides (see Table S1). According to Table 1 and Table S2 (see Supplementary information) at low nitrogen content ($x = 0.10$), the amount of N is not enough to form ideal CrN, therefore $Cr_{0.69}N_{0.31}$ with nitrogen deficiency was considered as a competing nitrogen-containing phase while the other two competing phases are considered to stay in their nitrogen-free form: fcc Fe-Ni alloy and fcc Co. At $x = 0.20$, the competing phases are the ideal CrN, and nitrogen-free Fe-Ni and Co. At higher nitrogen content, $x = 0.25$ and $x = 0.35$, the CrN phase cannot accommodate any more N atoms, therefore the surplus nitrogen atoms will occupy the interstitial position in the Fe-Ni-phase (See Table 1). Finally, at the maximum nitrogen content, $x = 0.50$, the competing phases are CrN, (Fe-Ni)N and CoN.

*Table 1. Compositions of (CrFeCoNi)$_{1-x}$N$_x$ solid solutions and three competing phases.*

| Nitrogen content $x$ in alloy | Competing phases | | |
|---|---|---|---|
| | **CrN** | **Fe-Ni-rich** | **Co-rich** |
| 0.00 | - | - | - |
| 0.10 | Cr$_{0.69}$N$_{0.31}$ | Fe-Ni | Co |
| 0.20 | CrN | Fe-Ni | Co |
| 0.25 | CrN | (Fe-Ni)$_{0.86}$N$_{0.14}$ | Co |
| 0.35 | CrN | (Fe-Ni)$_{0.67}$N$_{0.33}$ | Co |
| 0.50 | CrN | (Fe-Ni)N | CoN |

Figure 1 shows the configurational and magnetic entropies of mixing of (CrFeCoNi)$_{1-x}$N$_x$ solid solutions estimated according to Eq.3 and Eqs. S1-S3 in Supplementary information. One should note that contribution to the entropy S$_{conf}$ from "nitrogen-vacancy" sublattice is smaller than the one from metal sublattice, $S_{conf}^N < S_{conf}^{Me}$. Therefore, when nitrogen content is increased, the $S_{conf}$ becomes lower. At $x = 0.50$ the entropy reaches the minimum value, $S_{conf} = 0.5 \cdot S_{conf}^{Me}$, i.e., the solid solution becomes half-ordered. As for the competing phases, CrN has zero configurational entropy except when there is nitrogen deficiency in CrN. The Co-rich phase was considered as pure Co, therefore its configurational entropy is zero. In the Fe-Ni-rich phase there are contributions from metal and nitrogen sublattices. The latter is present when N content in solid solution $x > 0.20$. However, the entropy of Fe-Ni-rich phase is much smaller compared to one of (CrFeCoNi)$_{1-x}$N$_x$ solid solution. Therefore, the resulting configurational entropy of mixing is large enough, $\Delta S_{conf} \approx 0.87$ R (7.5 eV/ K), which should have a significant stabilization effect at finite temperatures.

To determine the free energy of mixing for paramagnetic alloys we also calculated the magnetic contribution to entropy $S_{mag}$ while considering different magnetic states for reference phases. One should note that for the magnetically ordered structures $S_{mag} = 0$ . For the ferromagnetic (CrFeCoNi)$_{1-x}$N$_x$ alloys, the magnetic states of the reference phases are antiferromagnetic CrN with orthorhombic structure, ferromagnetic Fe-Ni and Co; and for the paramagnetic alloys, the references are paramagnetic CrN, ferromagnetic Co and ferromagnetic Fe-Ni at $T = 500$ K or paramagnetic Fe-Ni at $T = 1000$ K. The magnetic entropy of mixing $\Delta S_{mag}$ of paramagnetic (CrFeCoNi)$_{1-x}$N$_x$ solid solution is smaller than $\Delta S_{conf}$ as it is shown in Figure 1. The $\Delta S_{mag}$ has a maximum at low N concentration and gradually decreases with increased N in (CrFeCoNi)$_{1-x}$N$_x$.

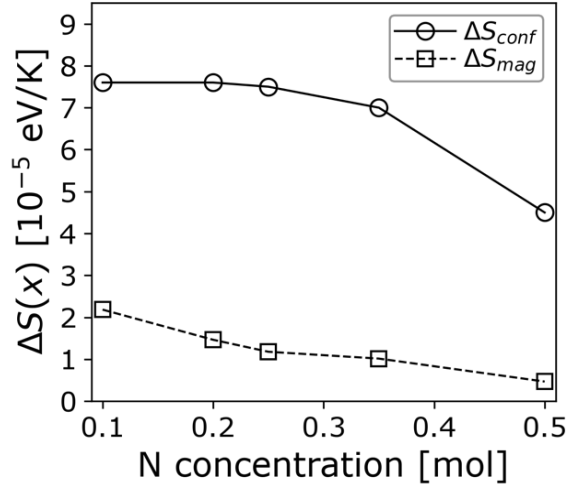

*Fig. 1. Mixing entropy of (CrFeCoNi)₁₋ₓNₓ solid solutions. The mixing entropy was defined with respect to three reference phases according to Eq.3.*

Figure 2 shows the mixing free energy $\Delta G$ of (CrFeCoNi)$_{1-x}$N$_x$ solid solutions as a function of N concentration. For both ferromagnetic and paramagnetic cases, the $\Delta G$ of (CrFeCoNi)$_{1-x}$N$_x$ solid solutions demonstrate similar trends, although the $\Delta G$ of the ferromagnetic solid solutions is higher. At finite temperatures, the $\Delta G$ becomes lower because of entropy factor, therefore one can expect a stabilization of single-phase solid solution with temperature. The (CrFeCoNi)$_{1-x}$N$_x$ solid solutions with $x = 0.20$ - $0.25$ demonstrate highest $\Delta G$, suggesting that this composition most likely decomposes into three reference phases. At lower and higher N concentrations, the mixing energy becomes lower, i.e., the tendency to segregation becomes weaker.

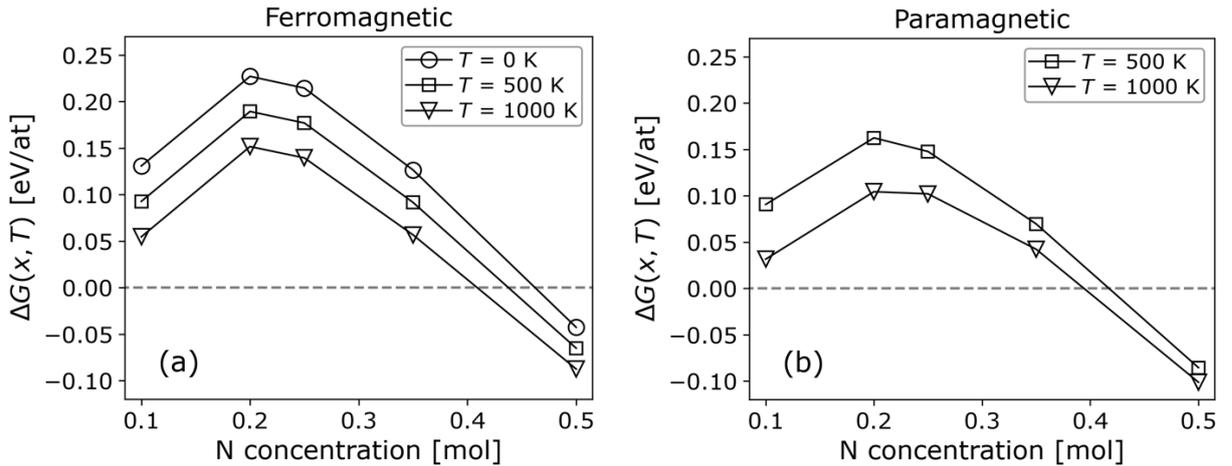

*Figure. 2. Mixing free energy of (CrFeCoNi)₁₋ₓNₓ solid solutions at (a) ferromagnetic and (b) paramagnetic states. The reference states for the ferromagnetic alloys are antiferromagnetic CrN, ferromagnetic Fe-Ni and Co; and for the paramagnetic alloys, the references are paramagnetic CrN, ferromagnetic Co and ferromagnetic Fe-Ni at T = 500 K or paramagnetic Fe-Ni at T = 1000 K. The compositions of reference phases are shown in Table 1.*

## 3.2 Experimental results

Cantor alloy-based thin films were grown by magnetron sputtering at different deposition temperatures and with different nitrogen contents. The detailed composition of both the metallic and nitrogen-containing films acquired from a combination of ERDA and PIXE can be found in the supplementary information (Table S5 and S6). The nitrogen-containing films will be referred to as $Me_{1-x}N_x$ (Me + N = 1 and with $0.0 \leq x \leq 0.5$).

### 3.2.1 Metallic films ($Me_{1.0}N_{0.0}$)

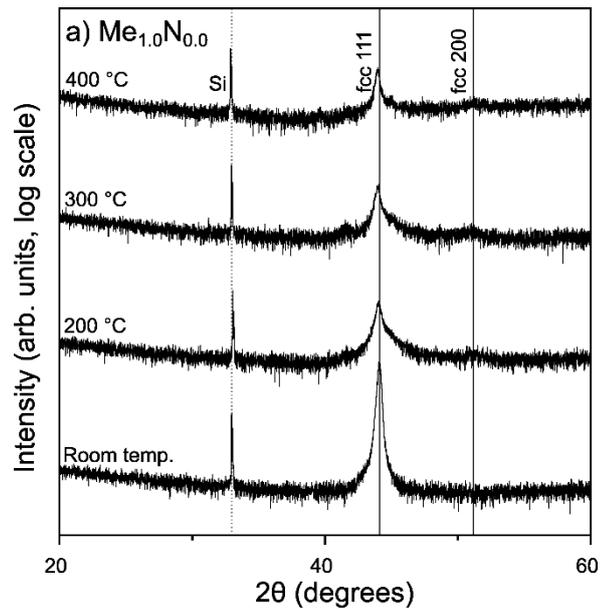

*Figure 3. X-ray diffractograms of CrFeCoNi ($Me_{1.0}N_{0.0}$) films grown at different deposition temperature from room temperature to 400 °C*

Figure 3 shows the X-ray diffractograms of the metallic films grown at different deposition temperatures starting from room temperature (RT) (i.e., without intentional heating), 200 °C, 300 °C, and 400 °C. The film grown at room temperature is fcc-structured with a (111) texture and a lattice parameter of 3.55 Å ± 0.02 Å, comparable to the Cantor alloy (lattice parameter = 3.6 Å). For the films deposited at higher temperature, the intensity of the fcc (111) peak is reduced and shows asymmetric broadening. The decrease in intensity and the broadening suggest the presence of at least two peaks partially overlapping. This could be an indication of a secondary orientation or phase [30]. More information on the metallic films can be found in section 2 of the supplementary information.

### 3.2.2 Nitrogen-containing films (Me$_{1-x}$N$_x$)

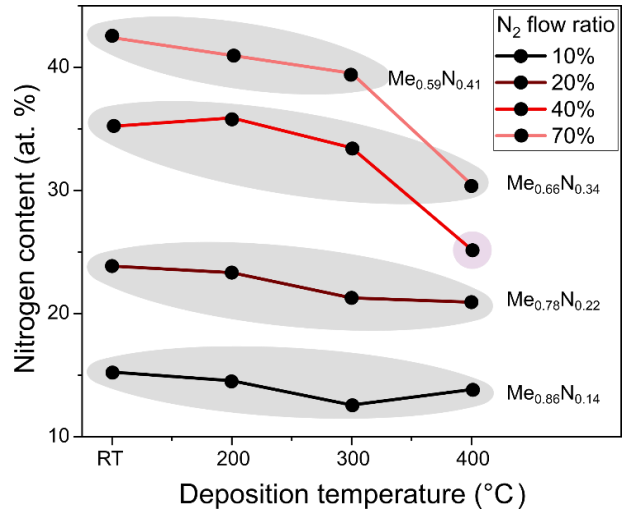

*Figure 1. Variation in nitrogen content of Me$_{1-x}$N$_x$ films for films deposited at different temperature, from room temperature (RT) to 400 °C. The composition is an average based on the nitrogen contents of the films contained in a group (indicated in gray). An approximate error bar of ± 1 at. % can be expected for each individual point.*

Four samples were grown for each deposition temperature at different relative nitrogen flow $F_N$ (10 – 70 %). Detailed compositions are provided as supplementary information, Table S6. Figure 4 shows the nitrogen content as a function of the deposition temperature for different N$_2$ flow ratio. For all films, the nitrogen content increased as the nitrogen flow increased from 10 to 70 %. For films deposited at room temperature and 200 °C, the nitrogen content does not significantly change, however, at higher temperature, the nitrogen content in the film decreased. Hereafter, the samples are referred according to N composition and categorized into four groups: Me$_{0.86}$N$_{0.14}$, Me$_{0.76}$N$_{0.22}$, Me$_{0.66}$N$_{0.34}$, and Me$_{0.59}$N$_{0.41}$, as indicated in Figure 4. The film grown at 400 °C and $F_N$ = 40 % is not included as the composition falls in between two groups and the microstructure, crystal structure, and mechanical properties of this film did not differ much from the film grown at $F_N$ = 70 % at the same temperature. More information on this film can be found in the supplementary information (Fig. S4).

The nitride-formation enthalpy of each of the metals (Cr, Fe, Co, and Ni) plays a large role in phase formation. Among these four metals, Cr has the lowest nitride-formation enthalpy [31], which means that nitrogen would preferentially bond with Cr among the available elements.

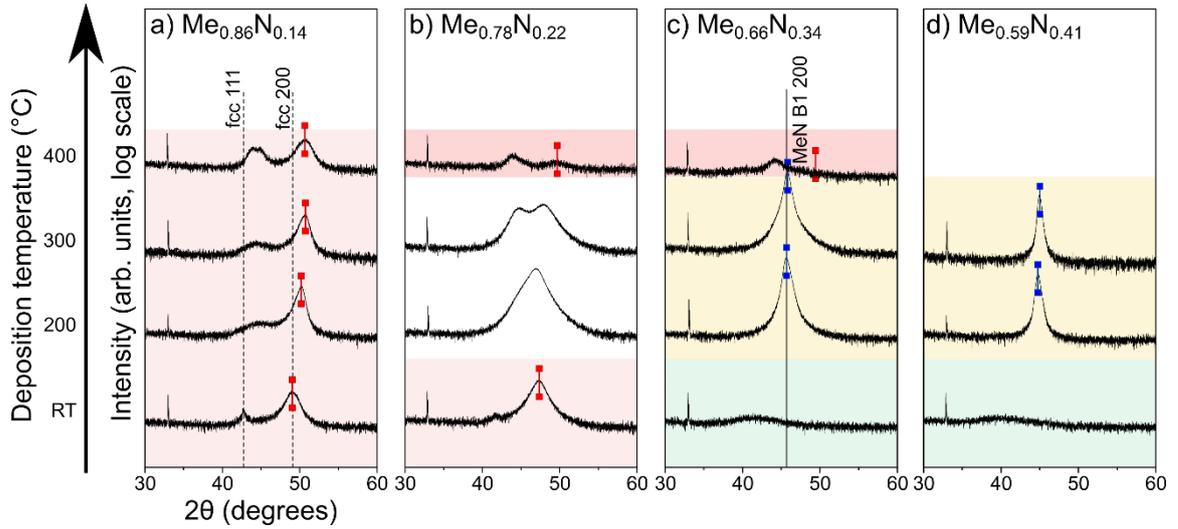

*Figure 2. XRD patterns of CrFeCoNi films with increasing amounts of nitrogen (x-axis) and grown at increasing deposition temperatures (y-axis). Dotted lines in (a) correspond to the fcc 111 and 200 reflections. Peaks marked with red marker correspond to the nitrogen-supersaturated fcc 200 reflections while peaks marked with blue marker correspond to the NaCl-MeN B1 200 reflections.*

Figure 5 shows X-ray diffractograms of the $Me_{1-x}N_x$ films where x increases from 0.14 to 0.41 and where each color represents a particular crystal structure. The region shaded in red [low N content/all temperature (T)] in figure 5 indicates the films where the stabilization of the fcc is observed ($x = 0.14$). The resulting crystal structure at room temperature is a nitrogen-supersaturated fcc structure with a larger lattice parameter (~3.87 Å) in comparison to the Cantor fcc (3.6 Å). The X-ray diffractograms of these films also indicate a change in texture to 200 in comparison to the 111 textured metallic film. When the deposition temperature is higher, a shift in the XRD peaks to higher angles is observed which implies a decrease of the lattice parameter to approximately 3.65 Å. At the highest deposition temperature of 400 °C, the X-ray diffractograms show the presence of an additional peak at an approximate $2\theta$ value of 45°, which could belong to a secondary phase. This nitrogen supersaturated fcc structure is retained up to a nitrogen content of $x = 0.22$ (Fig. 5b). The film grown at room temperature exhibits a larger lattice parameter (3.96 Å). The region shaded in dark red (intermediate N content/high T) in 5b and 5c represents the films which also crystallize in the metallic fcc structure despite the higher percentage of nitrogen in the plasma during growth. The higher deposition temperature may be a reason for the reduced nitrogen contents in the film in comparison to the other films belonging to the $Me_{0.66}N_{0.34}$ series. The X-ray diffractograms of the films with $x > 0.22$ indicate a stabilization of CrN isotype (Fig 5c and 5d, region shaded in yellow (high N content/intermediate T). These films are in a nearly stochiometric nitride phase with a 200 texture. This structure will be referred to as MeN B1. The lattice parameter of the MeN B1 was calculated to be $4.16 \pm 0.02$ Å. More information on these films can be found in figure S5 in the supplementary information

The diffractograms shaded with green (high N content/RT) on figure 5c and 5d indicate the amorphous films. These films were grown at room temperature with the highest amounts of nitrogen ($x = 0.34$ and 0.41). This region indicates the limit in solubility of nitrogen into the metallic fcc lattice.

From the X-ray diffractograms presented in figure 5b, the unshaded region (intermediate N content/intermediate T) represents the films $Me_{0.78}N_{0.22}$ grown at 200 °C and 300 °C. The XRD patterns of these films show complex features with broad asymmetric peaks where several peaks of diffraction are expected indicating towards the presence of multiple phases. In order to further investigate the phase evolution with nitrogen addition, TEM analysis was carried out on the films grown at 300 °C.

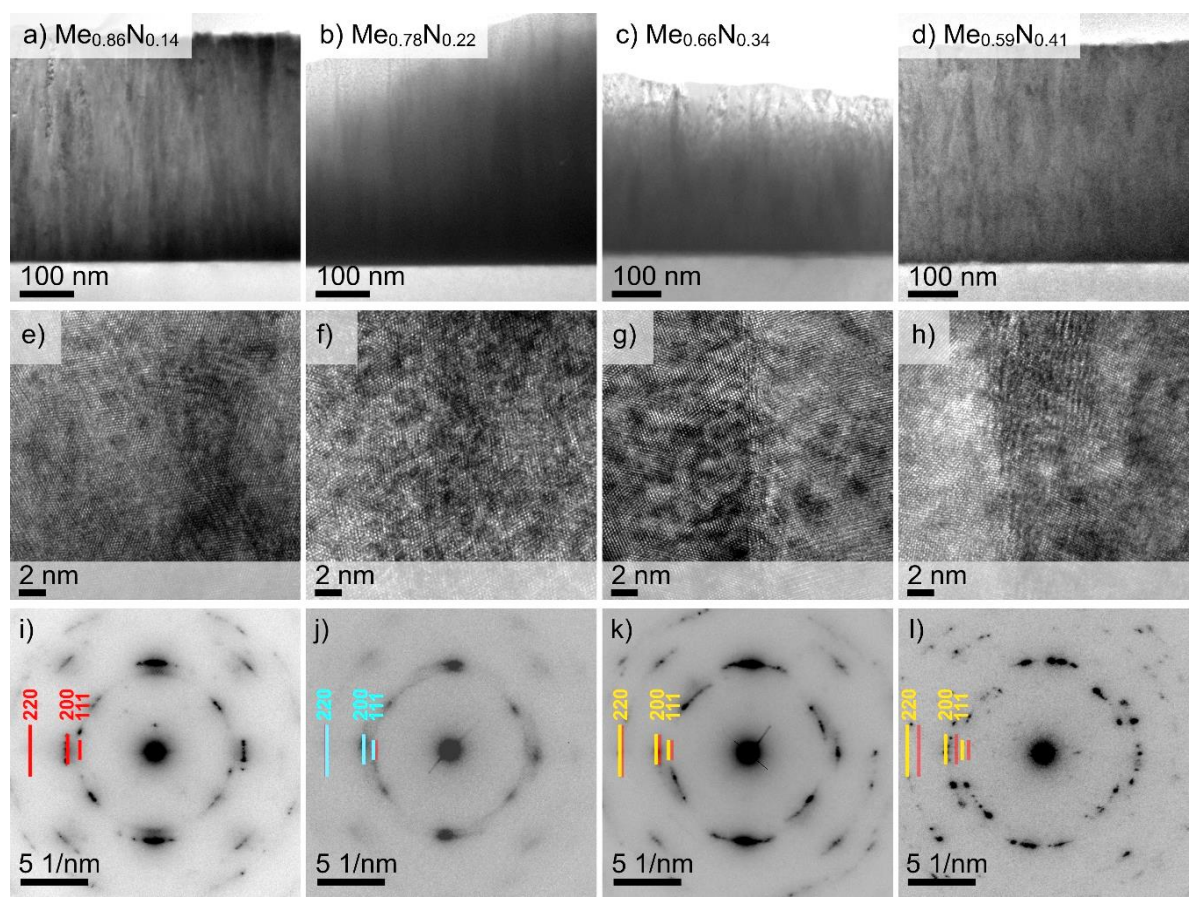

*Figure 6. (a-d) BF-TEM cross section images of nitrogen containing films deposited at 300 °C, corresponding HRTEM (e-h) and SAED patterns (i-l). Red marker lines in 6i indicate the position of reflections from the (111), (200) and (220) planes of the fcc structure. The blue and yellow markers in figure 6j, 6k and 6l indicated the positions of the indexed plane reflections while the red markers indicated the position of the (111), (200) and (220) plane reflections of the nitrogen supersaturated fcc as depicted in 6i.*

Figure 6 shows the BF-TEM cross sections of the films grow at 300 °C along with their corresponding HRTEM and SAED patterns. The BF-TEM images show a similar columnar structure for all the nitrogen-containing films (Figs. 6a - 6d). The SAED patterns of all films shown in figure 6i – 6l can be indexed to a cubic structure with increasing lattice parameters which suggests the expansion of the crystal lattice with the incorporation of nitrogen. This is also observed in the X-ray diffractograms where a shift in the peaks to smaller angles in observed (Fig. S3). The films are preferentially oriented along the 200 and misoriented in the in-plane direction in the case of $Me_{0.86}N_{0.14}$. As the nitrogen content is increased the preferential orientation is lost and the grains are more randomly oriented as seen for $Me_{0.58}N_{0.41}$.

The SAED pattern of the $Me_{0.78}N_{0.22}$ (Fig. 6j) indicates a single-phase solid solution, but the X-ray diffractogram along with STEM-EELS elemental mapping the suggests otherwise (Fig 5b and Fig. 7). This inconstancy is because segregation occurs at a nanoscale while the SAED patterns are acquired at a micron scale.

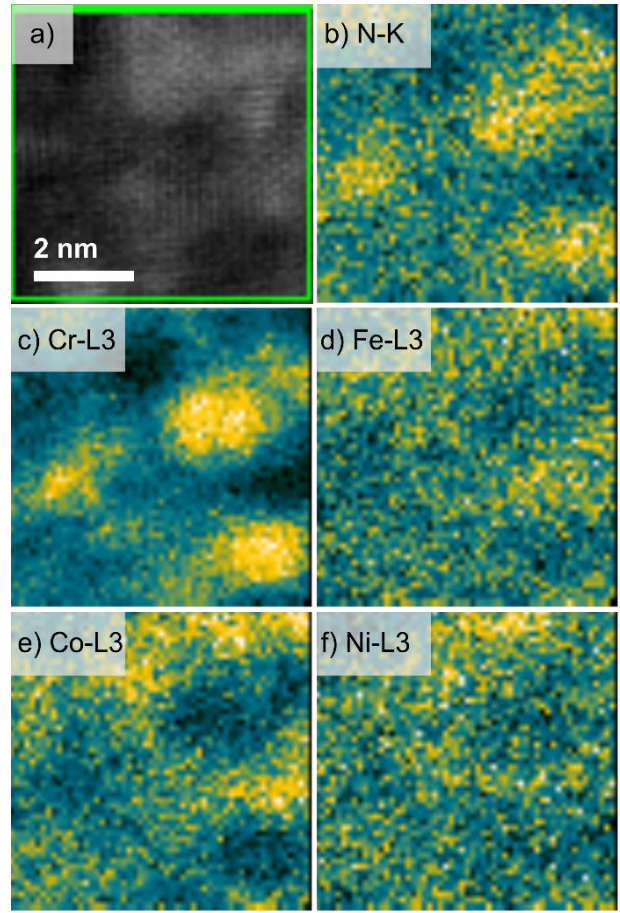

*Figure 7. (a) EELS spectrum image of $Me_{0.78}N_{0.22}$ grown at 300 °C along with corresponding elemental maps(b-f).*

The EELS elemental maps show that the composition is inhomogenous. The nitrogen signal is detected together with chromium, while cobalt, iron and nickel accumulate in adjacent locations.

Theoretical estimations of the free energy (see Fig. 2) show that, in both the ferromagnetic and the paramagnetic cases, the $\Delta G$ plots show similar trends with the maximum at $x = 0.20$-0.25, although the $\Delta G$ of the ferromagnetic solid solutions are higher. The experimentally observed segregation in $Me_{0.78}N_{0.22}$ films agree well with theory where the solid solutions with x = 0.20 - 0.25 demonstrate highest $\Delta G$. These observations also suggest that the samples with this composition decompose into the three reference phases. In $Me_{0.78}N_{0.22}$ films deposited at room temperature; segregation can presumably be constrained due to low diffusion rates. At intermediate deposition temperatures, 200 °C and 300 °C, diffusion is more significant, and the thermodynamic factor (high $\Delta G$) enables segregation in the $Me_{0.78}N_{0.22}$ films. Finally, at 400 °C, the entropy factor $\Delta S_{mix}$ can stabilize a single-phase solid solution structure in $Me_{0.78}N_{0.22}$ films.

At $x = 0.10$, the mixing free energy is lower, but still positive. This is because the CrN-reference phase is deficient with nitrogen (see Table 1), subsequently making it energetically less favorable than the ideal CrN phase. As a result, in solid solutions with $x < 0.20$ one can expect a stabilization of single-phase structure. At higher nitrogen concentrations, $x > 0.20$, the solid solutions also demonstrate lower $\Delta G$ (see Fig.2). This can be related to the fact that, according to our assumption on the segregation trend, the CrN phase cannot take more nitrogen and the remaining nitrogen atoms are taken by Fe-Ni- and Co-rich reference phases. This, in turn, can be energetically less favorable because the Fe-N, Ni-N and Co-N bonds are weaker than Cr-N bond. As a result, at higher N concentrations the $\Delta G$ becomes lower and one should expect stabilization of a single-phase structure in $Me_{1-x}N_x$ solid solutions. Note that, in Fig. 2, we are only analyzing the $\Delta G$ trends to qualitatively explain phase formation in experimental films, without making any direct comparison between theoretical and experimental transition temperatures (when $\Delta G$ becomes zero). There are two reasons for this: (i) samples with $x = 0.14$ and $x = 0.34$ of N which do not decompose, and which have $\Delta G > 0$ (see Fig. 2) can still be metastable single-phase solid solutions; (ii) chemical compositions of experimental $Me_{1-x}N_x$ films are not equimolar while the $\Delta G$ calculations were performed for equimolar composition.

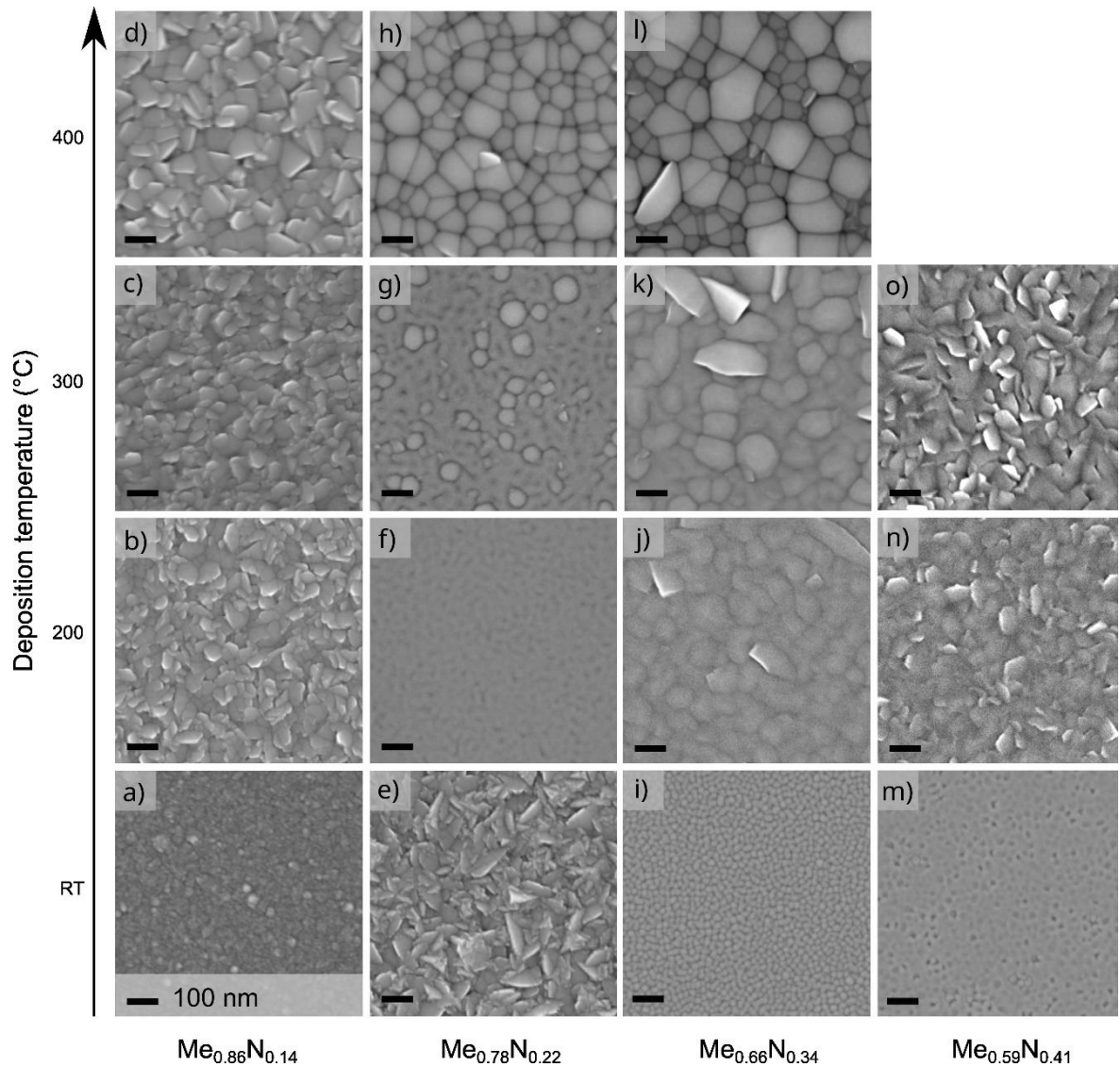

*Figure 8. Top-view SEM images of nitrogen-containing films showing the change in morphology for different nitrogen content and deposition temperature.*

Some features of the multiple phases can also be found in surface morphology. Figure 8 shows top-view SEM images of the nitrogen-containing films with the deposition temperature increasing along the y-axis and nitrogen content increasing along the x-axis. Focusing on the films grown at 300 °C, the nitrogen supersaturated $Me_{0.86}N_{0.14}$ film shows a flake-like morphology (Fig. 8c). As the nitrogen content is increased, a large change in morphology is observed where the nucleation of the secondary phase can be observed, consistent with the X-ray diffractogram of $Me_{0.78}N_{0.22}$ (Fig 8g). As the nitrogen content is further increased and the MeN B1 is stabilized, the films display large well-crystallized grains (Figs. 8k and 8o). The 200-texture observed in the X-ray diffractograms of $Me_{0.66}N_{0.34}$ is also visible in the SEM image of the film (Fig. 8k).

*Table 2. Hardness and reduced Young's modulus values of nitrogen-containing films obtained from nanoindentation in load-controlled mode.*

| Film | Deposition temperature (°C) | Structure | Hardness (± 2 GPa) | Reduced Young's modulus (± 10 GPa) | Resistivity (± 5 μΩ.cm) |
|---|---|---|---|---|---|
| $Me_{0.86}N_{0.14}$ | RT | fcc | 15.5 | 294 | 156 |
| | 200 | fcc | 17.0 | 238 | 160 |
| | 300 | fcc | 15.8 | 238 | 140 |
| | 400 | fcc | 14.9 | 238 | 118 |
| $Me_{0.78}N_{0.22}$ | RT | fcc | 14.9 | 284 | 159 |
| | 200 | Multi-phase | 27.8 | 284 | 140 |
| | 300 | Multi-phase | 21.9 | 277 | 127 |
| | 400 | fcc | 14.1 | 233 | 115 |
| $Me_{0.66}N_{0.34}$ | RT | Amorphous | 16.0 | 267 | 177 |
| | 200 | MeN B1 | 22.1 | 301 | 178 |
| | 300 | MeN B1 | 20.7 | 270 | 168 |
| | 400 | fcc | 14.0 | 251 | 202 |
| $Me_{0.59}N_{0.41}$ | RT | Amorphous | 15.6 | 248 | 194 |
| | 200 | MeN B1 | 22.2 | 309 | 196 |
| | 300 | MeN B1 | 25.6 | 272 | 196 |

Table 2 shows the hardness and reduced Young's modulus of the nitrogen-containing MeN-series thin films. Films with the same crystal structures display similar hardness and modulus values, such that all the films stabilized as an fcc phase show hardness near $15.0 \pm 2.0$ GPa. The highest hardness was observed in the multi-phase samples belonging to the $Me_{0.78}N_{0.22}$ group of films. The increase in hardness is consistent with the nanocomposite-like structure of the film. Resistivity values of the films are also provided in table 2. The trends indicate that the resistivity is influenced largely by the nitrogen content in the films and less by the crystal structure itself unlike the hardness and modulus.

## 4. Conclusion

In summary, we performed a systematic investigation of phase formation in CrFeCoNi solid solutions with the addition of nitrogen. DFT calculations of mixing energy $\Delta G$ were carried out with reference to three competing phases states: CrN, Fe-Ni- and Co-rich. The mixing energy as a function of nitrogen concentration has a maximum at $x = 0.20$-$0.25$, which explains the segregation in experimental films with $x = 0.22$ of nitrogen. The lower values of $\Delta G$ at $x \leq 0.10$ and $x \geq 0.25$ concentration of nitrogen can qualitatively explain the stabilization of single-phase structures in $(CrFeCoNi)_{1-x}N_x$. The films undergo segregation when the nitrogen content in the film is sufficient to form CrN as reference phase. At $x = 0.20$, the competing reference phases are at their most stable states: ideal CrN and nitrogen-free Fe-Ni and Co. As a result, the mixing free energy has the highest value at $x = 0.20$. The configuration entropy contributes to stabilizing single-phase structures in $(CrFeCoNi)_{1-x}N_x$ at high temperatures. The theoretically and experimentally obtained results on phase formation well corroborate each other. This study provides a guideline for ceramic Cantor alloy-related systems and can assist researchers in the field when choosing deposition parameters to grow films with specified structures and properties.

## Acknowledgements

The work was supported financially by the VINNOVA Competence Centre FunMat-II (grant no. 2016-05156), the Swedish Government Strategic Research Area in Materials Science on Functional Materials at Linköping University (Faculty Grant SFO-Mat-LiU No. 2009 00971), the Knut and Alice Wallenberg foundation through the Wallenberg Academy Fellows program (KAW-2020.0196) and the Wallenberg Scholar Grant (KAW-2018.0194), The Swedish Foundation for Strategic research through the Future research leaders 6 program SSF-FFL 15-0290, and by the Swedish Research Council (VR) under project number 2021-03826 and 2019-05403. The computations were enabled by resources provided by the Swedish National Infrastructure for Computing (SNIC) located at National Super Computer Centre (NSC) in Linköping, partially funded by the Swedish Research Council through Grant Agreement No. 2018-05973. Accelerator operation at Uppsala University was supported by Swedish Research Council VR-RFI (Contract No. 2019-00191). The Swedish Research Council and the Swedish Foundation for Strategic Research are acknowledged for access to ARTEMI, the Swedish National Infrastructure in Advanced Electron Microscopy (Grant No. 2021-00171 and RIF21-0026).

## Author contributions

**Smita G. Rao:** Conceptualization, Experimental Investigation, Data curation, Formal analysis, Writing - original draft. **Boburjon Mukhamedov:** Conceptualization, Theoretical Investigation, Data curation, Formal analysis, Writing - original draft. **Gyula Nagy:** Experimental data curation, Formal analysis Writing - review & editing. **Eric N. Tseng:** Experimental data curation, Formal analysis, Writing - review & editing. **Rui Shu:** Formal analysis, Writing - review & editing. **Robert Boyd:** Experimental data curation, Formal analysis, Writing - review & editing. **Daniel Primetzhofer**: Experimental data curation, Writing - review & editing, Funding acquisition. **Per Persson:** Writing - review & editing, Funding acquisition. **Björn Alling:** Conceptualization, Theoretical Investigation, Supervision, Formal analysis, Writing - review & editing. **Igor A. Abrikosov:** Project administration, Conceptualization, Supervision, Formal analysis, Writing - review & editing, Funding acquisition. **Arnaud le Febvrier:** Conceptualization, Investigation, Supervision, Formal analysis, Writing - review & editing. **Per Eklund:** Project administration, Conceptualization, Supervision, Writing - review & editing, Funding acquisition.

## Data availability statement

Data are available from the corresponding author on reasonable request.

**Supplementary information for**

**Phase formation in CrFeCoNi nitride thin films**

Smita G. Rao[†*1], Boburjon Mukhamedov[†*2], Gyula Nagy[4], Eric N. Tseng[1], Rui Shu[1], Robert Boyd[3], Daniel Primetzhofer[4], Per O. Å. Persson[1], Björn Alling[2], Igor A. Abrikosov[2], Arnaud le Febvrier[1], Per Eklund[1]

*[1] Thin Film Physics Division, Department of Physics, Chemistry, and Biology (IFM), Linköping University, Linköping, Sweden-58183.*

*[2] Theoretical Physics Division, Department of Physics, Chemistry, and Biology (IFM), Linköping University, Linköping, Sweden-58183.*

*[3] Plasma & Coatings Physics Division, Department of Physics, Chemistry, and Biology (IFM), Linköping University, Linköping, Sweden-58183.*

*[4] Department of Physics and Astronomy, Uppsala University, Lägerhyddsvägen 1, S-75120 Uppsala, Sweden*

*[†]Corresponding author: e-mail address: smita.gangaprasad.rao@liu.se; boburjon.mukhamedov@liu.se*

## 1. Calculated structures and configurational entropy

*Table S1. Nitride formation energies taken from Materials Project [1–4].*

| Compound | Predicted formation energy (eV/atom) |
|----------|--------------------------------------|
| CrN | -0.663 |
| FeN | -0.315 |
| CoN | -0.074 |
| NiN | 0.253 |

[*] S.G.R and B. M. contributed equally to this study

*Table S2. Number of atoms in the calculated supercells for (CrFeCoNi)$_{1-x}$N$_x$ solid solutions and three competing phases.*

| N content, mol | (CrFeCoNi)N solid solution | | Competing phases | | | | | |
|---|---|---|---|---|---|---|---|---|
| | | | CrN-rich | | Fe-Ni-rich | | Co-rich | |
| | Me | N | Cr | N | Fe-Ni | N | Co | N |
| 0.00 | 108 | 0 | - | - | - | - | - | - |
| 0.10 | 108 | 12 | 108 | 48 | 108 | 0 | 108 | 0 |
| 0.20 | 108 | 27 | 108 | 108 | 108 | 0 | 108 | 0 |
| 0.25 | 108 | 36 | 108 | 108 | 108 | 18 | 108 | 0 |
| 0.35 | 108 | 54 | 108 | 108 | 108 | 54 | 108 | 0 |
| 0.50 | 108 | 108 | 108 | 108 | 108 | 108 | 108 | 108 |

Table S2 shows the number of atoms in the calculated SQS supercells for equiatomic (CrFeCoNi)$_{1-x}$N$_x$ solid solutions and three reference phases. The numbers of metal atoms in (CrFeCoNi)$_{1-x}$N$_x$ and three reference phases are equal. The concentration of N in (CrFeCoNi)$_{1-x}$N$_x$ changes from $x = 0.10$ up to 0.50. In (CrFeCoNi)$_{1-x}$N$_x$ supercell with $x = 0.10$, there are not enough N atoms to form ideal CrN reference phase (see Table S2). Therefore, we generated CrN with nitrogen deficiency and random occupation of octahedral positions in fcc Cr supercell. When $x > 0.20$ the amount of N atoms becomes more than CrN phase can take, therefore extra nitrogen atoms will occupy the interstitial positions in Fe-Ni phase. This assumption is made based on our experimental data that showed that the Co-composition anti-correlated with the N-composition. Finally, for (CrFeCoNi)$_{1-x}$N$_x$ solid solution with $x = 0.50$, the reference states are CrN, (Fe-Ni)N and CoN.

The entropy $S$ of the (CrFeCoNi)$_{1-x}$N$_x$ solid solutions and corresponding competing phases was calculated as a sum of configurational $S_{conf}$ and magnetic $S_{mag}$ entropies:

$$S = S_{conf} + S_{mag}$$

The configurational entropy in nitrogen containing solid solutions has two contributions: from metal sublattice and from nitrogen sublattice. These contributions should be accounted for according to the molar fractures between metal and nitrogen atoms:

$$S_{conf} = S_{conf}^{total} = x_{Me} \cdot S_{conf}^{Me} + x_N \cdot S_{conf}^{N} \qquad \textbf{(S1)}$$

where $x_{Me}$ and $x_N$ denote molar fractures of all metal atoms and nitrogen, respectively. Maximum value of $x_N$ is 0.5 and corresponds to ideal alloy nitride. $S_{conf}^{Me}$ and $S_{conf}^{N}$ are the configurational entropies in metal and nitrogen sublattices, respectfully, which are defined as:

$$S_{conf}^{Me} = -R \sum_i c_i \cdot ln c_i \qquad \textbf{(S2)}$$

$$S_{conf}^{N} = -R(c_N \cdot lnc_N + c_V \cdot lnc_V) \tag{S3}$$

where $R$ denote universal gas constant; $c_i$ denote concentration of $i$-th metal atoms; $c_N$ and $c_V$ respectively denote concentrations of nitrogen and vacancies in sublattice of nitrogen. Note, the maximum value of $c_N$ is 1.0, which corresponds to the ideal alloy nitride when all octahedral positions are occupied.

Calculated $S_{conf}$ and $S_{mag}$ of (CrFeCoNi)$_{1-x}$N$_x$ solid solutions and three competing phases are given in Tables S3 and S4, respectively.

*Table S3. Configurational entropies of metal $S_{conf}^{Me}$ and nitrogen $S_{conf}^{N}$ sublattices and corresponding total configurational entropies $S_{conf}^{total}$ of (CrFeCoNi)$_{1-x}$N$_x$ solid solutions and three competing phases.*

| $x_N$ (mol) | $S_{conf}$ of (CrFeCoNi)$_{1-x}$N$_x$ (10-5 eV/at) | | | $S_{conf}$ of competing phases (10-5 eV/at) | | | | | | | | |
|---|---|---|---|---|---|---|---|---|---|---|---|---|
| | | | | CrN-rich | | | Fe-Ni-rich | | | Co-rich | | |
| | $S_{conf}^{Me}$ | $S_{conf}^{N}$ | $S_{conf}^{total}$ | $S_{conf}^{Me}$ | $S_{conf}^{N}$ | $S_{conf}^{total}$ | $S_{conf}^{Me}$ | $S_{conf}^{N}$ | $S_{conf}^{total}$ | $S_{conf}^{Me}$ | $S_{conf}^{N}$ | $S_{conf}^{total}$ |
| 0.10 | 11.96 | 3.01 | 11.07 | 0.00 | 5.93 | 1.82 | 5.98 | 0.00 | 5.98 | 0.00 | 0.00 | 0.00 |
| 0.20 | 11.96 | 4.85 | 10.54 | 0.00 | 0.00 | 0.00 | 5.98 | 0.00 | 5.98 | 0.00 | 0.00 | 0.00 |
| 0.25 | 11.96 | 5.50 | 10.35 | 0.00 | 0.00 | 0.00 | 5.98 | 3.89 | 5.68 | 0.00 | 0.00 | 0.00 |
| 0.35 | 11.96 | 5.98 | 9.97 | 0.00 | 0.00 | 0.00 | 5.98 | 5.98 | 5.98 | 0.00 | 0.00 | 0.00 |
| 0.50 | 11.96 | 0.00 | 5.98 | 0.00 | 0.00 | 0.00 | 5.98 | 0.00 | 2.99 | 0.00 | 0.00 | 0.00 |

*Table S4. Magnetic entropy $S_{mag}$ of paramagnetic (CrFeCoNi)$_{1-x}$N$_x$ solid solutions and three competing phases. For magnetically ordered structures $S_{mag} = 0$.*

| $x_N$ (mol) | $S_{mag}$ of (CrFeCoNi)$_{1-x}$N$_x$ (10-5 eV/at) | $S_{mag}$ of Competing phases (10-5 eV/at) | | |
|---|---|---|---|---|
| | | CrN-rich | Fe-Ni-rich | Co-rich |
| 0.10 | 0.27 | 0.05 | 0.00 | 0.00 |
| 0.20 | 0.24 | 0.26 | 0.00 | 0.00 |
| 0.25 | 0.20 | 0.26 | 0.00 | 0.00 |
| 0.35 | 0.18 | 0.26 | 0.00 | 0.00 |
| 0.50 | 0.12 | 0.26 | 0.00 | 0.00 |

## 2. Metallic films (Me$_{1.0}$N$_{0.0}$)

*Table S5. Composition of metallic films estimated form ERDA and PIXE*

| Temperature (°C) | Atomic percentage (± 1 at. % ) | | | | |
|---|---|---|---|---|---|
| | Cr | Fe | Co | Ni | N |
| Room temp. | 22.7 | 29.7 | 27.1 | 20.6 | - |
| 200 | 20.2 | 29.8 | 27.5 | 22.4 | - |
| 300 | 20.9 | 30.0 | 27.5 | 21.7 | - |
| 400 | 20.2 | 31.0 | 27.7 | 21.2 | - |

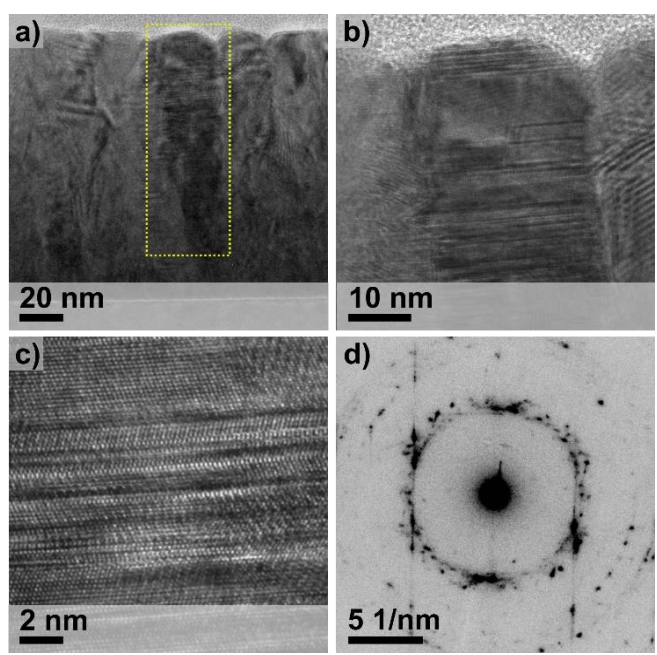

*Figure S1. a) TEM cross-section micrograph of metallic film grown at 300 °C. (b) and (c) HRTEM images showing the presence of stacking faults in the columnar grains. (d) SAED pattern taken from film cross-section*

Figure S1 shows the cross-section TEM images of the metallic film grown at a deposition temperature of 300 °C. The cross-section image reveals the columnar growth structure of the films. A contrast difference is observed in the columns which is an indication of the presence of defects in the columns. HRTEM images of the region marked in yellow are shown in figures S1 b and S1 c. These images clearly show the presence of stacking faults within the columns. The selected area diffraction pattern (SAED) taken from the film cross section is seen in figure S1 d. The polycrystalline nature of the film can be observed from the selected area diffraction pattern (SAED) with the presence of diffused spots and rings.

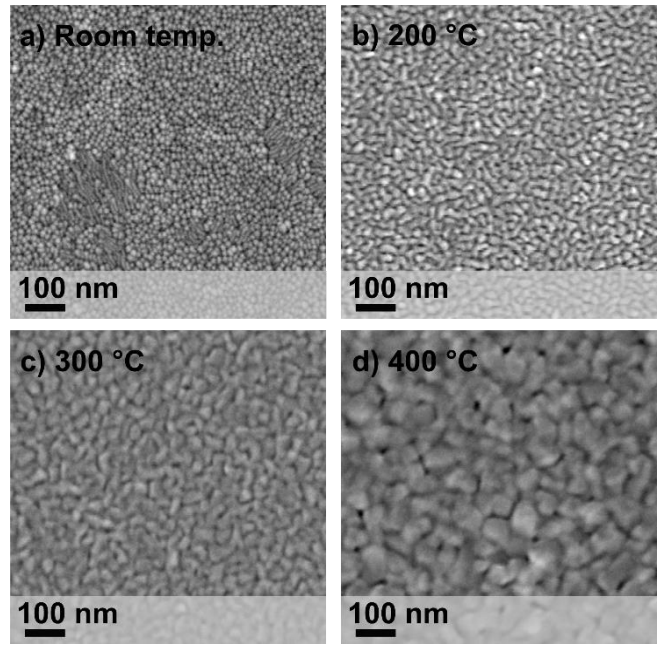

*Figure S2. SEM planar view of CrFeCoNi films grown at increasing deposition temperatures*

Figure S2 shows the surface view SEM micrographs of CrFeCoNi metallic films grown at increasing deposition temperatures. At room temperature (Fig. S2 a) the film exhibits a small and fine-grained structure. As the deposition temperature increases, coalescence and grain growth are further promoted resulting in larger sized grains.

## 3. Nitrogen containing films

4.

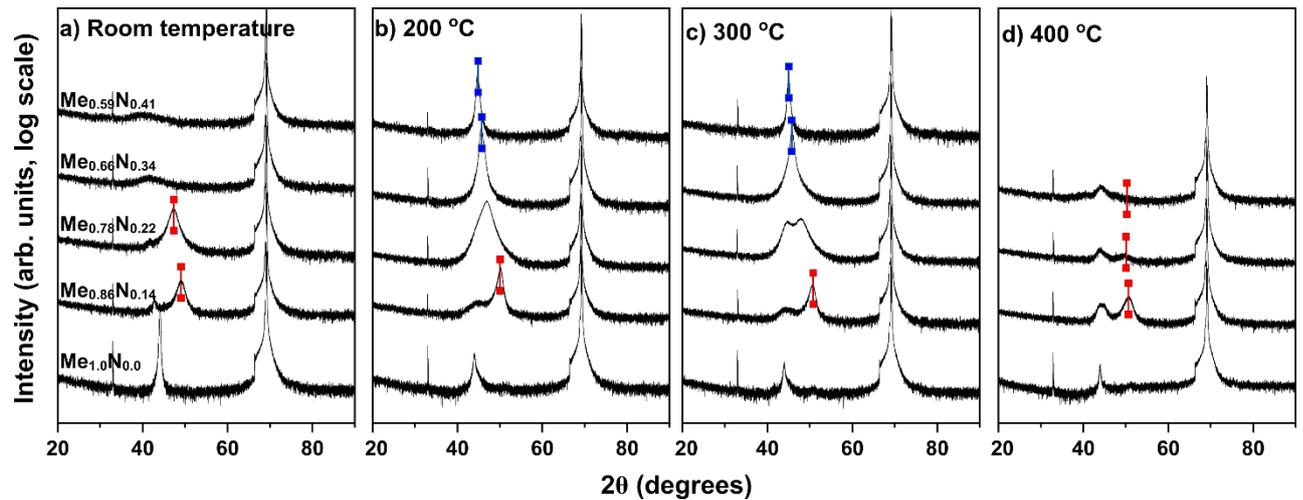

*Figure S3. XRD patterns of $Me_{1-y}N_y$ films grouped according to deposition temperature. Red markers indicate the fcc 200 reflections while the blue markers indicate the NaCl MeN B1 200 reflections.*

Figure S3 shows the XRD patterns of the $Me_{1-y}N_y$ films grouped according to the deposition temperature.

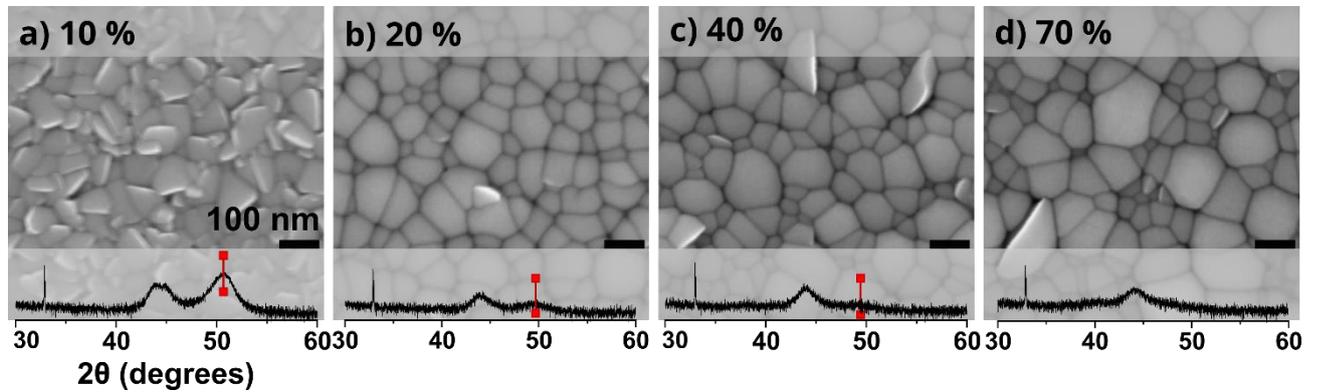

*Figure S4. planar SEM images of films grown at 400 °C while varying $F_N$ from 10 to 70 %.*

Figure S4 shows the morphology along with inset XRD patterns of the nitrogen containing films grown at 400 °C. At this deposition temperature the films show no significant change in terms of morphology and crystal structure. The hardness of both films grown with a nitrogen flow ratio corresponding to 40 and 70 % was found to be ≈ 13 GPa.

*Table S6: Composition and thickness of CrFeCoNi films estimated from ToF-ERDA, RBS and PIXE. Me+N =1, where Me is Cr+Fe+Co+Ni*

| Nitrogen % in total gas | Temperature (°C) | Atomic percentage (1.5 - 3 at. %) | | | | | N/Me | Thickness (at/cm²) |
|---|---|---|---|---|---|---|---|---|
| | | Cr | Fe | Co | Ni | N | | |
| 10 % | Room temp. | 17.16 | 25.32 | 23.27 | 18.99 | 15.26 | 0.18 | 3780 |
| | 200 | 16.59 | 26.17 | 23.48 | 19.20 | 14.56 | 0.17 | 3770 |
| | 300 | 17.54 | 26.29 | 23.84 | 19.75 | 12.58 | 0.14 | 4030 |
| | 400 | 17.19 | 26.14 | 23.72 | 19.11 | 13.84 | 0.16 | 3950 |
| 20 % | Room temp. | 15.23 | 23.59 | 20.23 | 17.08 | 23.87 | 0.31 | 4840 |
| | 200 | 15.54 | 23.78 | 20.24 | 17.11 | 23.33 | 0.30 | 4580 |
| | 300 | 15.29 | 23.71 | 21.07 | 18.65 | 21.28 | 0.27 | 4900 |
| | 400 | 15.98 | 23.20 | 21.61 | 18.28 | 20.93 | 0.26 | 4730 |
| 40 % | Room temp. | 12.48 | 20.42 | 17.37 | 14.51 | 35.22 | 0.54 | 4780 |
| | 200 | 12.40 | 20.56 | 17.13 | 14.01 | 35.90 | 0.56 | 4730 |
| | 300 | 12.70 | 21.75 | 17.29 | 14.79 | 33.47 | 0.50 | 4950 |
| | 400 | 13.94 | 22.50 | 18.25 | 14.83 | 30.48 | 0.44 | 4250 |
| 70 % | Room temp. | 11.90 | 19.04 | 14.62 | 12.01 | 42.42 | 0.74 | 5160 |
| | 200 | 11.70 | 19.29 | 15.28 | 12.76 | 40.97 | 0.69 | 4680 |
| | 300 | 12.33 | 19.27 | 15.34 | 13.55 | 39.52 | 0.65 | 4720 |

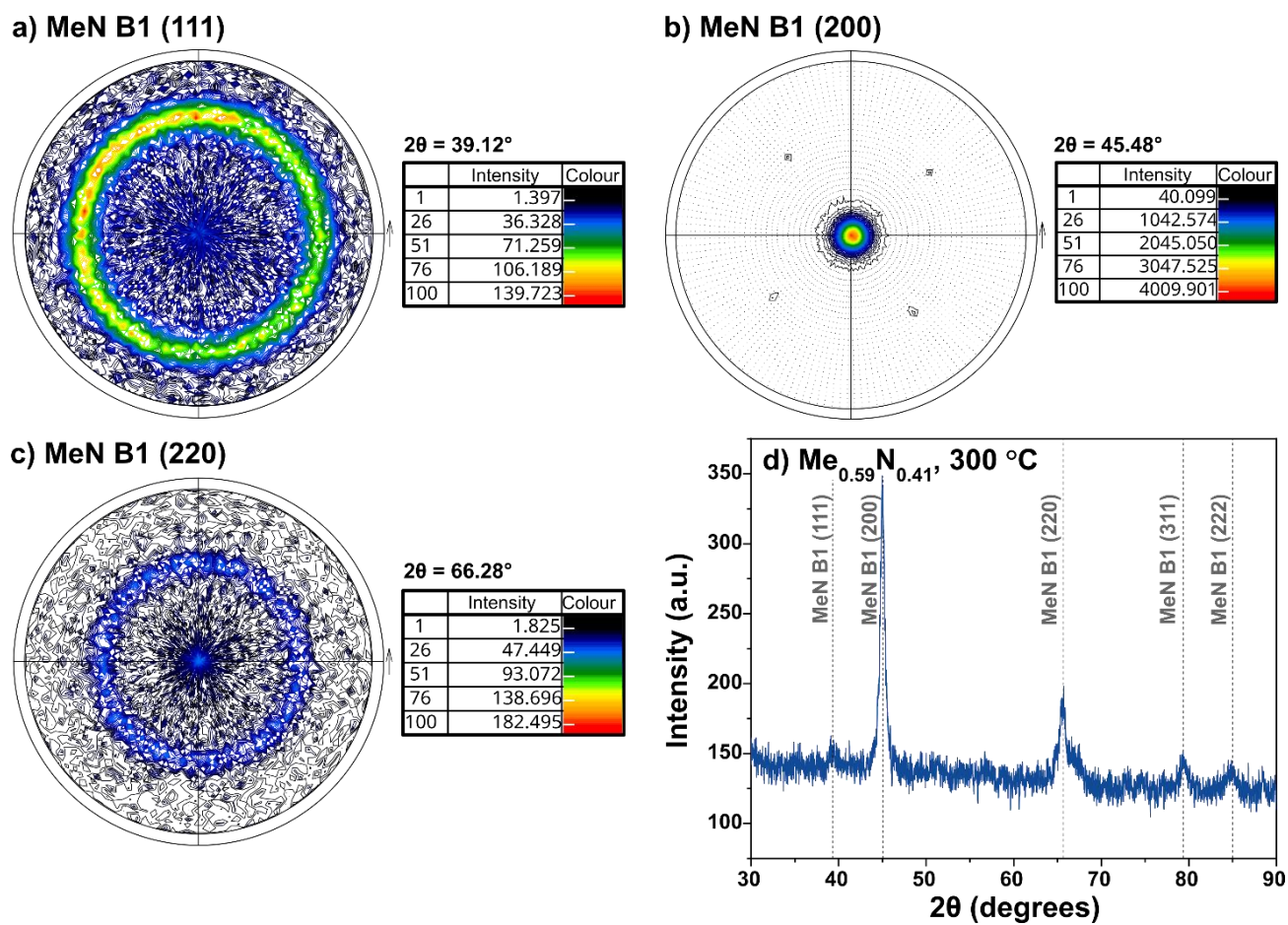

*Figure S5. (a-c) Pole figures of $Me_{0.59}N_{0.41}$ grown at 300 °C. d) GIXRD pattern showing all peaks corresponding to the NaCl MeN B1 structure.*

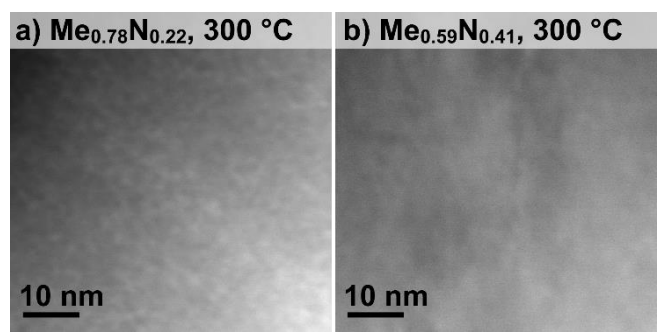

*Figure S6. Cross section STEM images of $Me_{0.78}N_{0.22}$ and $Me_{0.59}N_{0.41}$ grown at 300 °C, depicting the level of segregation in the films.*

Figure S6 shows STEM cross-section images of the understoichiometric, decomposed film (S6 a) and nearly stoichiometric single-phase film (S6 b). Brighter contrast corresponds to heavier elements while the darker contrast indicates the lighter elements.